\documentclass[useAMS,usenatbib,twocolumn]{mn2e}

\usepackage{dcolumn}
\usepackage{graphicx}
\usepackage{url}
\usepackage{color}

\newcommand{\cm}{cm$^{-1}$}

\newcommand{\ai}{\textit{ab initio}}

\newcommand{\eqref}[1]{(\ref{#1})}

\newcommand{\duo}{{\sc Duo}}

\title[ExoMol: IX The spectrum of AlO]{ExoMol molecular line lists: IX The spectrum of AlO}

\date{\today}
\author[A.T. Patrascu, S.N. Yurchenko and J. Tennyson ]{\large Andrei T.
Patrascu,
Sergei N. Yurchenko and Jonathan Tennyson \\
Department of Physics and Astronomy, University College London, London WC1E 6BT,
UK}
\date{Accepted XXXX. Received XXXX; in original form XXXX}

\pagerange{\pageref{firstpage}--\pageref{lastpage}} \pubyear{2012}
\begin{document}
\maketitle
\begin{abstract}
  Accurate line lists are calculated for aluminium monoxide covering the pure
  rotation, rotation-vibration and electronic (B -- X blue-green and A
  -- X infrared bands) spectrum.  Line lists are presented for the main
  isotopologue, $^{27}$Al$^{16}$O, as well as for $^{27}$Al$^{17}$O,
  $^{27}$Al$^{18}$O and $^{26}$Al$^{16}$O. These line lists are suitable for
  high temperatures (up to 8000 K) including those relevant to exoplanetary
  atmospheres and cool stars. A combination of empirical and \textit{ab
  initio} methods is used: the potential energy curves were previously
  determined to high accuracy by fitting to extensive data from analysis of
  laboratory spectra; a high quality {\it ab initio} dipole moment curve is
  calculated using quadruple zeta basis set and the multi-reference
  configuration interaction (MRCI) method. Partition functions plus full line
  lists of transitions are made available in an electronic form as
  supplementary data to this article and at \url{www.exomol.com}.

\end{abstract}
\begin{keywords}
molecular data; opacity; astronomical data bases: miscellaneous; planets and
satellites: atmospheres; stars: low-mass
\end{keywords}

\label{firstpage}

\section{Introduction}

Aluminium monoxide (AlO) is an interesting astronomical species whose spectrum is prominent
in a new class of Nova-stars first discovered by
\citet{09TeZixx.AlO}
 of which the
most prominent examples are probably V838 Mon and V4332 Sgr
\citep{62MeKeDe.AlO,84BeGrxx.AlO,jt357, 05TyCrGo.AlO,12BaVaMa.AlO}. These two
objects defined a new type of eruptive variables called intermediate luminosity
red transients and the observational data showed the intense presence of the
near-infrared A -- X system of the AlO radical. Indeed this  A -- X
band is also found to be fairly prominent in a variety of cool, oxygen rich
stars \citep{84BeGrxx.AlO,12BaVaMa.AlO}: besides the Mira variables discussed
above AlO emissions were also observed in the OH/IR stars and two bright
infrared sources \citep{12BaVaMa.AlO}.

Transitions in the blue-green B -- X system have been observed in sunspots
\citep{13SrViSh.AlO} and  the red supergiant VY Canis Majoris \citep{13KaScMe.AlO},
in which millimeter-wave rotational transitions have also been observed
\citep{09TeZixx.AlO}. Finally, AlO spectra have been used to try and determine
abundance of the long-lived, radioactive $^{26}$Al isotope
\citep{04BaAsLa.AlO}.

Terrestrially AlO emissions arise from rocket exhausts in the atmosphere
\citep{65Joxxxx.AlO,96KnPiMu.AlO}. Its spectrum is also extensively used  in
the laboratory to monitor  AlO in plasmas and other applications
\citep{95BeMoUr.AlO,99NaCoxx.AlO,01GlSeKr.AlO,03ZhLixx.AlO,14BaMoLe.AlO,14SuPa.AlO}.

These  applications, combined with technological uses of AlO spectra, have
motivated a number of laboratory studies which have produced molecular
constants characterising the lowest three states of AlO $X{}^{2}\Sigma^+$,
A\,$^{2}\Pi$ and B\,$^{2}\Sigma^+$. There have also been attempts to produce
line lists. \citet{11PaHoxx.AlO} constructed a comprehensive line list for the
X -- B system for temperatures up to 6000~K but did not provide a
transition dipole, so all their transition intensities are only relative.
\cite{11LaBexx.AlO} performed a combined analysis of the A --X   and B --
X band systems involving 21~500 lines; we compare with some of their results
below. There is, however, no single line list that combines a comprehensive set
of transition frequencies with an accurate model for the transition
intensities. It is this that we aim to do here as part of the ExoMol project.
ExoMol  aims to provide line lists of spectroscopic transitions for key
molecular species which are likely to be important in the atmospheres of
extrasolar planets and cool stars; its aims, scope and methodology have been
summarised by \citet{jt528}. Line lists for $^2\Sigma^+$ XH molecules, X = Be,
Mg, Ca, have already been published \citep{jt529}, as well as for a number of
closed-shell diatomics \citep{jt563,jt583,jt590}.  In the present paper, we
present rotation-vibration transition lists and associated spectra for AlO.
These line lists are particulary comprehensive and should be valid for
temperatures up to 8000 K.

\section{Method}

Rotation-vibration line lists for the three lower electronic
states of AlO were obtained by direct solution of the nuclear motion
 Schr\"{o}dinger equation using program \duo\ \citep{jt606}.
The calculations require both a potential energy curve (PEC) for each of the
three states considered and also couplings between these curves. These curves
were taken from our previous study \citep{jt589}, which computed {\it ab
initio} potential energies, spin-orbit and electronic angular momenta
couplings, and refined them using available experimental data. Nuclear motion
calculations using these refined curves showed that the observed transition
frequencies and energy levels could be reproduced with root mean square error
of only 0.07 cm$^{-1}$. In order to cover all vibrational excitations below
35,000~\cm, we have increased the sizes of the vibrational basis set to 90 for
each of the X, A, and B states from those used by \citet{jt589}. The
ranges of rotational excitations are listed in Table~\ref{tab:lifetime}.

\subsection{Dipole moments}
There appears to be no experimental measurements of any AlO transition dipoles.
For this reason we constructed  new dipole moment curves (DMC) using high level
{\it ab initio} calculations. These are compared to previous,  high-level {\it
ab initio} determinations \citep{99ZeBlCh.AlO} below. The \ai\ calculations
were performed using MOLPRO \citep{molpro.method}; we used multi-reference
configuration interaction (MRCI) methods with different choices of basis sets.
Our optimal basis choice was aug-cc-pVQZ; the active space used in MOLPRO
representation was (9,4,4). Electronic dipole moments as function of
bondlength, $R$, were computed as the expectation value
\begin{equation}
\mu(R) = e \langle \Psi_M | \sum_i \underline{r}_i | \Psi_N \rangle ,
\end{equation}
where the integral and the summation run over the electron coordinates, denoted
by $\underline{r}_i$, and $e$ is the charge of the electron.  For permanent
dipole moments, the electronic wavefunctions in the bra and ket are the same,
ie $M = N$, and the dipole moment, which is denoted $\mu(M)$ below, also
contains a term due to permanent nuclear charge. For transition
dipole moments, $M \neq N$, and the dipole is denoted is $\mu(M - N)$ below.
For transition dipole moments, care must be taken to ensure the that the dipole
phases are consistent as a function of $R$ \citep{jt573,jt589}.

Our calculations produce the values for the dipole at equilibrium given in
Table ~\ref{tab:dipol} which compare well to the previous results obtained by
\citet{99ZeBlCh.AlO}.  Our ground state value of the dipole and that of
\citet{99ZeBlCh.AlO} are both slightly smaller than the value 4.60 D used in
the JPL database \citep{jpl} which was taken from the earlier calculations of
\citet{82LeLixx.AlO}.

\begin{table}
\caption{{\it Ab initio} electric dipole and transition dipole moments in Debye
at $R=1.76$ \AA.}
\label{tab:dipol}
\begin{center}
\begin{tabular}{lrr}
\hline\hline
Transition moments                           &  This Work          &
\citet{99ZeBlCh.AlO}$^{a}$     \\
\hline
X\,$^{2}\Sigma^+$                   &  $-$4.39 & $-$4.24        \\
A\,$^{2}\Pi$                        &  $-$1.30 & $-$1.40  \\
B\,$^{2}\Sigma^+$                   &  $-$2.18 & $-$2.27    \\
X\,$^{2}\Sigma^+$ --B\,$^{2}\Sigma^+$ &   1.85 &  1.66    \\
X\,$^{2}\Sigma^+$ --A\,$^{2}\Pi$    &     0.61 &  0.61   \\
B\,$^{2}\Sigma^+$ --A\,$^{2}\Pi$    &  $-$0.046              \\
\hline
\end{tabular}
\\
$^{a}$ The signs of the diagonal dipoles have been changed to conform
to the convention used by MOLPRO.
\end{center}
\end{table}

\begin{figure}
\vspace{1cm}
\begin{center}
\includegraphics[width=220pt]{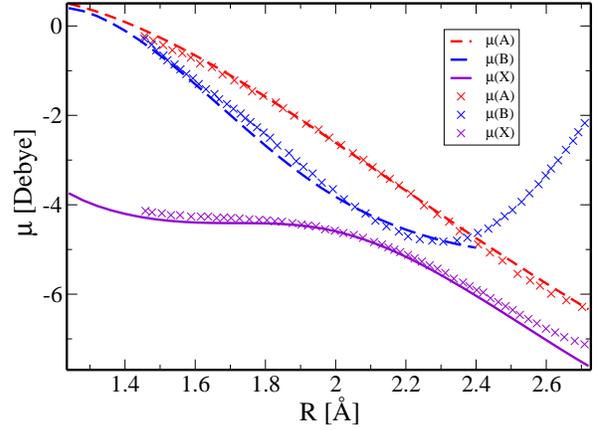}
\caption{{\it Ab initio} permanent dipole moment curves for AlO for the lowest three electronic
states. The previous calculations by
\citet{99ZeBlCh.AlO} are represented by crosses.}
\label{fig:dip1}
\end{center}
\end{figure}

\begin{figure}
\vspace{1cm}
\begin{center}
\includegraphics[width=220pt]{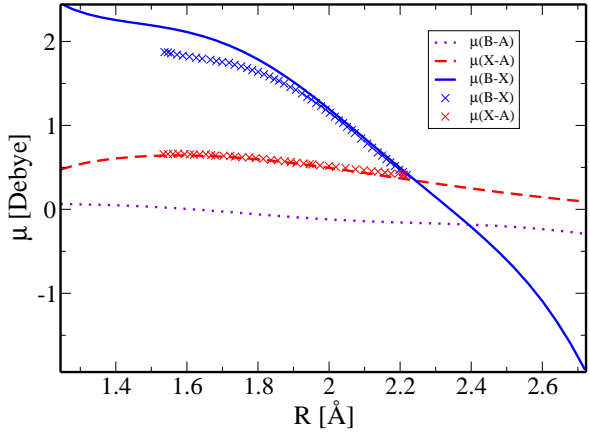}
\caption{{\it Ab initio} transition dipole moment curves for AlO linking the lowest three electronic
states. The previous calculations by
\citet{99ZeBlCh.AlO} are represented by crosses.}
\label{fig:dip2}
\end{center}
\end{figure}

 Figures~\ref{fig:dip1} and ~\ref{fig:dip2} compare our
calculated diagonal and off-diagonal DMCs, respectively, with those of
\citet{99ZeBlCh.AlO}. The agreement is good. Our calculations suggest that the
$\mu$(B-A) DMC is small at all geometries meaning that the B -- A band
will be very weak; a similar conclusion was reached by \citet{83PaLaLe.AlO}.

The \ai\ DMC grid points were used directly in \duo\ to produce a line list for
AlO.

There is a lack of experimental data on AlO transition dipoles or transition
intensities. Table~\ref{tab:lifetime} therefore compares the lifetime for the
B\,$^{2}\Sigma^+$ state with experimental data available from
\citet{72JoCaBr.AlO,75DaCrZa.AlO}    and two {\it ab initio} estimates from
\citet{83PaLaLe.AlO}. For \citet{83PaLaLe.AlO} we have taken their figures
which include the small contribution from the weak B -- A decay channel
since this contribution is also included in our estimate. Our lifetimes were
computed by summing over all decays from a given B\,$^{2}\Sigma^+$ $(v,J)$. Our
results in Table~\ref{tab:lifetime} are for $J=0.5$; calculations for $J=24.5$,
which lies in the region of the band head, give lifetimes about 0.5~\%\ longer.
We conclude that the lifetimes are not strongly $J$-dependent. In common with
the other studies we find that the lifetime grows slowly with $v$. Our results
are intermediate between the two predictions of  \citet{83PaLaLe.AlO} and
slightly shorter than, but marginally consistent with, the measurements of
\citet{75DaCrZa.AlO}. The measurements of \citet{72JoCaBr.AlO} give longer
lifetimes than all studies and we suggest these are too long.

\begin{table*}
\caption{Radiative lifetimes (nsec) for B\,$^{2}\Sigma^+$ state of AlO, compared
to the measurements of \citet{72JoCaBr.AlO} and \citet{75DaCrZa.AlO}, and the
two separate calculations
of  \citet{83PaLaLe.AlO}. }
\label{tab:lifetime}
\begin{center}
\begin{tabular}{lccccc}
\hline\hline
Vibrational Level& This work& \citet{83PaLaLe.AlO} I& \citet{83PaLaLe.AlO} II& \citet{72JoCaBr.AlO}& \citet{75DaCrZa.AlO} \\
\hline
0		& 92.4	& 88.1& 109.9 & 128 $\pm$ 6& 100 $\pm$ 7 \\
1		& 94.5	& 90.5& 112.6 & 125 $\pm$ 3& 102 $\pm$ 7 \\
2		& 96.7	& 93.0& 115.2 & 130 $\pm$ 7& 102 $\pm$ 4 \\
\hline
\end{tabular}
\end{center}
\end{table*}

\begin{figure}
\begin{center}
\includegraphics[width=220pt]{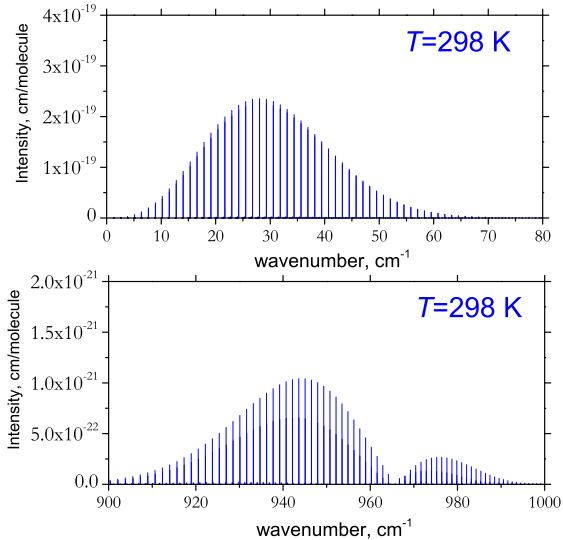}
\caption{Computed spectra of $^{27}$Al$^{16}$O at $T$=298 K  given as sticks with
the intensity (cm molecule$^{-1}$) represented by their height.
Upper panel: rotational region; lower panel: vibrational fundamental.}
\label{fig:vibf}
\end{center}
\end{figure}

\subsection{Partition function}

Partition functions for AlO were calculated by summing all the calculated
energy levels below using \duo\ \citep{jt606}. When summing these levels it is
necessary to multiply by the appropriate degeneracy factors.  Since we follow
HITRAN \citep{03FiGaGo.partfunc} and use the full nuclear spin degeneracy, the
degeneracy factor, $g$, is given by $(2J+1)(2I_{\rm Al}+1)(2I_{\rm O}+1)$ where
$J$ is the total angular momentum quantum number obtained by adding the
rotational and spin angular momenta. $I_{\rm Al}$ and $I_{\rm O}$ are the
nuclear spins of the isotopes of Al and O in the given isotopologue. Explicit
inclusion of these nuclear spin factors accounts for hyperfine effects which we
make no attempt to resolve. These factors  are 11, 6, 1, 6 and 1 for $^{26}$Al,
$^{27}$Al, $^{16}$O, $^{17}$O and $^{18}$O, respectively.

Table~\ref{tab:finallevels} compares our results for $^{27}$Al$^{16}$O
with those of \citet{84SaTaxx.partfunc}.
We have multiplied the results of \citet{84SaTaxx.partfunc} by the
appropriate nuclear spin factors to bring their results into line
with our convention outlined above.
  Table~\ref{tab:finallevels} shows good agreement
between our $^{27}$Al$^{16}$O partition function and that given by
\citep{84SaTaxx.partfunc} at temperatures above 1000 K for which their
results are valid. At lower temperatures we also agree well with the
partition function given by JPL \citep{jpl} who, for example, give
$Q(300) = 3926.45$ which is slightly lower than our value of 3966.90,
probably due to neglect of the contribution of excited vibrational
states.

As we use all ro-vibrational energy levels there are no issues with convergence
of this sum. We follow \citet{jt263} and represent our partition function using
the following functional form
\begin{equation}
 \log_{10}{Q(T)}=\sum_{n=0}^8 a_{n} [\log_{10}T]^{n}
\label{eq:part}\end{equation} where the fitting parameters $a_{n}$ are given in
Table~\ref{tab:part_isotopo}. These fits reproduce the partition functions for the
entire region below $9000$~K with a relative root-mean-square (rms) errors of
better than 1.6~\%.

\begin{table}
\caption{Partition function, $Q(T)$, for $^{27}$Al$^{16}$O, as a function of temperature.}
\label{tab:finallevels}
\begin{center}
\begin{tabular}{lrrrr}
\hline\hline
T / K  &	Q(T) & T & Q(T) &  \citet{84SaTaxx.partfunc}   \\
\hline

10	&	134.73	&	1000&  17595.69&  17693.4	 \\
20	&	265.35	&	2000&  57302.56&  57060.6	 \\
30	&	396.01	&	3000& 138649.47&  135168	 \\
40	&	526.68	&	4000& 283031.72&  274740	 \\
50	&	657.37	&	5000& 508066.61&  487708 \\
60	&	788.07	&	6000& 828224.93&  793980	 \\
70	&	918.78	&	7000&1254881.70& 1216536	 \\
80	&	1049.50	&	8000&1795616.84& 1781718	 \\
100	&	1310.96 &	    &			&		\\
200	&	2621.45 &	    &			&		\\
300	& 	3966.90 &	    &			&		\\
400	&	5407.40 &	    &			&		\\	
500	&	6988.27 &	    &			&		\\
600	&	8734.58 &	    &			&		\\
750	&	11692.87&	    &			&		\\
\hline
\end{tabular}
\end{center}
\end{table}

\begin{table*}
\caption{Partition function parameters for various isotopologues, see Eq.~(\ref{eq:part})}
\label{tab:part_isotopo}
\begin{center}
\begin{tabular}{lrrrr}
\hline\hline
	&	$^{27}$Al$^{16}$O	&	$^{27}$Al$^{18}$O	&	$^{26}$Al$^{16}$O	&	$^{27}$Al$^{17}$O	\\
\hline									\\
$a_0$   &        -1.04093681038&        -1.59727184392&        -0.677569296333&        -0.541673874468\\
$a_1$   &         9.64080670554&         11.9848644854&         9.21909668136&         10.8157290731\\
$a_2$   &        -14.3512337912&        -18.2945471026&        -13.6457740234&        -16.3230021925\\
$a_3$   &         13.0627960677&         16.7055086746&         12.4154930773&         14.8789580623\\
$a_4$   &        -7.20103828655&        -9.21814194734&        -6.84559240132&        -8.20306751046\\
$a_5$   &         2.49431683028&         3.17902318889&         2.37492955521&         2.83289571369\\
$a_6$   &        -0.538890191754&        -0.677958762340&        -0.514966111414&        -0.607255974396\\
$a_7$   &         0.0673568508718&         0.0828183310157&         0.0647424233238&         0.0749009373167\\
$a_8$   &        -0.00372906050126&        -0.004450320873&        -0.00360977781061&        -0.00407761291349\\
\hline
\end{tabular}
\end{center}
\end{table*}

\begin{table*}
\caption{Summary of our AlO linelists.}
\label{tab:summarized}
\begin{center}
\begin{tabular}{lrrrr}
\hline\hline
		& $^{27}$Al$^{16}$O  	& $^{27}$Al$^{18}$O 	& $^{26}$Al$^{16}$O 	& $^{27}$Al$^{17}$O \\
\hline
X\,$^{2}\Sigma^+$	&			&			&		      &			\\
Maximum $v$ 	& 66			& 69	  		& 66		  & 68\\
Maximum $J$ 	& 300.5			& 300.5 		& 300.5		  & 300.5\\
A\,$^{2}\Pi$	&			&			&		          &	\\
Maximum $v$	& 63			& 65			& 62		  & 64 \\
Maximum $J$	& 300.5			& 300.5			& 300.5		  & 300.5\\
B\,$^{2}\Sigma^+$	&			&			&		      &	\\								
Maximum $v$	& 40			& 41			& 39		  & 40\\
Maximum $J$	& 232.5			& 241.5			& 230.5		  & 237.5  \\
\\
Number of lines	& 4\,945\,580		& 5\,365\,592		& 4\,866\,540 & 5\,148\,996\\
\hline
\end{tabular}
\end{center}
\end{table*}

\subsection{Line list calculations}

Line lists were calculated for the four isotopologues $^{27}$Al$^{16}$O,
$^{27}$Al$^{18}$O, $^{27}$Al$^{17}$O, and $^{26}$Al$^{16}$O. All
rotation-vibration states were considered and transitions satisfying the dipole
selection rule $\Delta J = 0, \pm 1$. These line lists span frequencies up to
35~000 cm$^{-1}$ ($\lambda > 0.286~\mu$m). The procedure described above was used
to produce line lists, i.e. catalogues of transition frequencies
$\tilde{\nu}_{ij}$ and Einstein coefficients $A_{ij}$, for four Aluminium oxide
isotopologues $^{27}$Al$^{16}$O, $^{27}$Al$^{18}$O, $^{27}$Al$^{17}$O, and
$^{26}$Al$^{16}$O. The full line list for each of the studied isotopologues are
summarised in Table~\ref{tab:summarized}.

\section{Results}

The line lists contain about 5 million transitions each and, therefore, for
compactness and ease of use, are divided into separate energy level and
transitions file. This is done using standard ExoMol format \citep{jt548} which
is based on a method originally developed for the BT2 line list \citep{jt378}.
Extracts for the start of the $^{26}$Al$^{16}$O files are given in
Tables~\ref{tab:levels} and \ref{tab:trans}.  The full line list for each of
these isotopologues can be downloaded from the CDS, via
\url{ftp://cdsarc.u-strasbg.fr/pub/cats/J/MNRAS/xxx/yy}, or
\url{http://cdsarc.u-strasbg.fr/viz-bin/qcat?J/MNRAS//xxx/yy}.  The line lists
and partition function together with auxiliary data including the potential
parameters and dipole moment functions, as well as the absorption spectrum
given in cross section format \citep{jt542}, can all be obtained from there as
well as at \url{www.exomol.com}.

\begin{figure}
\begin{center}
\includegraphics[width=220pt]{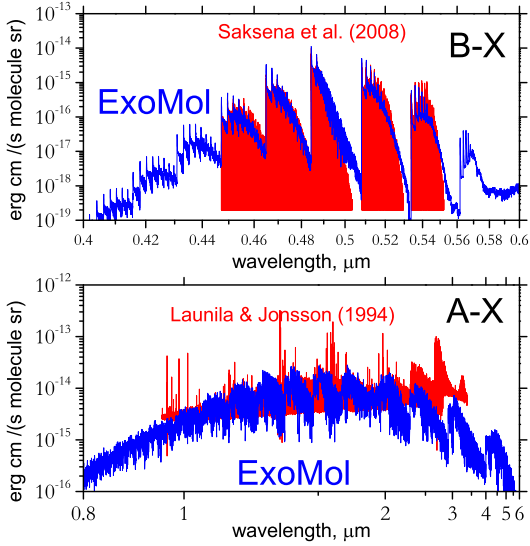}
\caption{Overview of the theoretical (ExoMol) and experimental
\citep{94LaJoxx.AlO,08SaDeSu.AlO} spectra of  $^{27}$Al$^{16}$O. The
theoretical spectra were obtained as cross sections convolved with a Gaussian
line profile of width 1~cm$^{-1}$ assuming the local thermal equilibrium at
$T=2000$~K. The experimental A -- X and B -- X spectra were scaled by 1$\times 10^{-17}$ and  5$\times 10^{-18}$, respectively.}
\label{fig:3000IR}
\end{center}
\end{figure}

Figure \ref{fig:vibf}  shows the rotational component, and the P- and weaker
R-branches of the vibrational fundamental ($v=0 - 1$) obtained at $T$=298~K. As
has been noted before \citep{82LeLixx.AlO}, the X-state dipole is very flat
in the equilibrium region. As the strength of a $\Delta v =1$
vibration-rotation transition depends on the slope of the dipole in this
region, this  causes the vibrational fundamental to be particularly weak.
Therefore this feature, which lies between 10 and 11 $\mu$m, is unlikely to be
astronomically important. Our calculations suggest that the overtones ($\Delta
v > 1$) are also weak so the entire AlO vibration-rotation spectrum is unlikely
to feature strongly in astronomical objects.

\begin{figure}
\begin{center}
\includegraphics[width=220pt]{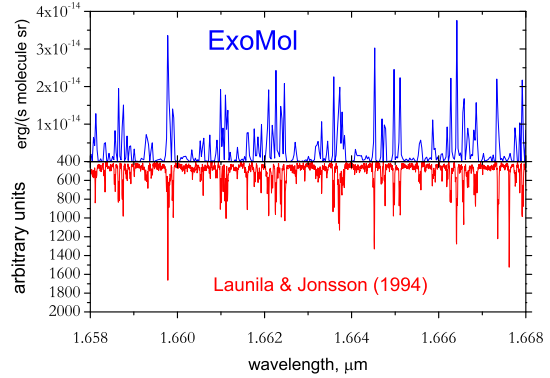}
\caption{A -- X emission spectrum of $^{27}$Al $^{16}$O,
comparison with experiments of
\citet{94LaJoxx.AlO} at $T$=3200~K. The
theoretical spectrum was obtained as cross sections convolved with a Doppler
line profile assuming the local thermal equilibrium at $T=3200$~K. }
\label{fig:comp}
\end{center}
\end{figure}

\begin{figure}
\begin{center}
\includegraphics[width=220pt]{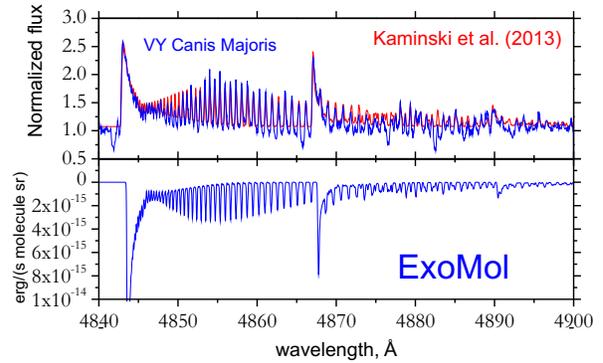}
\caption{B -- X $\Delta v =0$ emission spectrum at $T_{\rm rot}$=700~K  and $T_{\rm vib}$=2200~K
compared with an astronomical
spectrum obtained by \citet{13KaScMe.AlO}
for VY Canis Majoris ($T_{\rm rot}$  and $T_{\rm vib}$  in \citep{13KaScMe.AlO}).
Cross sections (lower part) were obtained by convolving with a Gaussian profile of width 0.3 cm$^{-1}$.}
\label{fig:spec4}
\end{center}
\end{figure}

Much more significant at infrared wavelengths is the A -- X electronic
band. Figure~\ref{fig:3000IR} shows an overview of the A -- X and B --
X electronic transitions which are presented as absorption spectra generated
at $T=2000$~K. Our spectra are compared to available experimental data
\citep{94LaJoxx.AlO,08SaDeSu.AlO}: we note that these measurements do not give
absolute intensities, so have been scaled by us. A more detailed comparison of
a portion of the  A -- X spectrum with the experimental results of
\citet{11LaBexx.AlO} is presented in Fig. ~\ref{fig:comp}. Figure
~\ref{fig:spec4} compares the theoretical spectrum obtained here with an
astronomical spectrum of \citet{13KaScMe.AlO}. The agreement is remarkable. The
calculation performed using the vibrational $T_{\rm vib}$ and rotational
$T_{\rm rot}$ temperatures of 2200~K and 700~K, respectively as suggested by
\citet{13KaScMe.AlO}. We use a gaussian convolution with the half-width at
half-maximum derived of 0.3~\cm\ to match the spectrum by \citet{13KaScMe.AlO}.
Also shown (in red) is the simulation of their observed spectrum by
\citet{13KaScMe.AlO}; for this they generated their own line list based on the
line positions of \citet{08SaDeSu.AlO}, transition moments of
\citet{99ZeBlCh.AlO}, Franck-Condon factors of \citet{85CoNaxx.AlO} and
rotational line-strength factors which they computed themselves. We note that
our line list provides all these data within a single framework and without
making any underlying assumptions about the Franck-Condon approximation or
rotational form factors.

\begin{figure}
\begin{center}
\includegraphics[width=220pt]{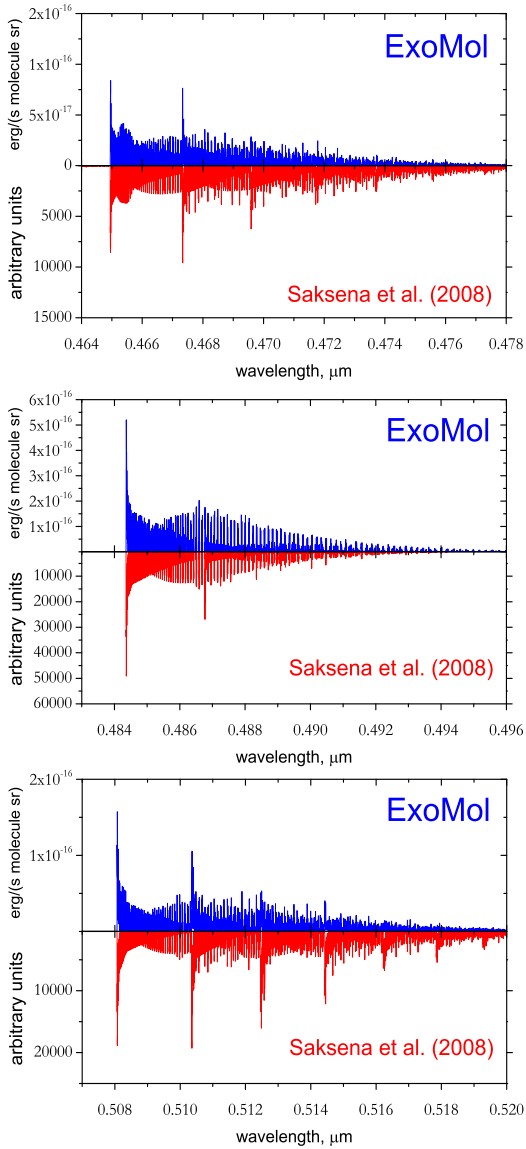}
\caption{Emission spectra of three sub-bands within the X -- B band at 1700~K
compared to the experiment of \citet{08SaDeSu.AlO} (up). Top panel:
$\Delta v=1$, middle panel: $\Delta v=0$, lower panel: $\Delta v=-1$.
The experimental data is in
arbitrary units; calculated cross sections
were obtained by convolving with a Doppler profile at 1700~K. }
\label{fig:spec5}
\end{center}
\end{figure}

\begin{figure}
\begin{center}
\includegraphics[width=220pt]{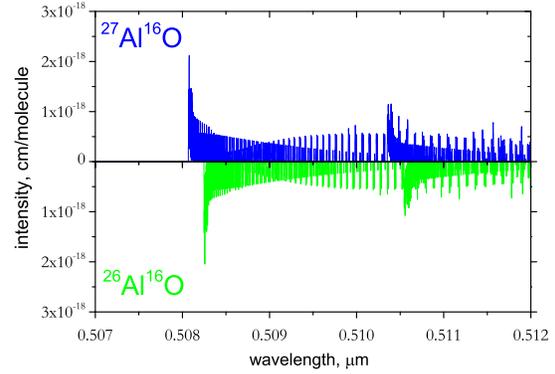}
\caption{Calculated B -- X  $v'-v'' = 0 - 1$
absorption spectrum at 1700~K for $^{26}$Al$^{16}$O and $^{27}$Al$^{16}$O
obtained by convolving with a Doppler profile at 1700~K.}
\label{fig:spec2c}
\end{center}
\end{figure}

Figure~\ref{fig:spec5} compares the B -- X emission spectrum obtained in
this work with accurate experimental results of \citet{08SaDeSu.AlO}. Again the
agreement is very good. Finally, Fig.~\ref{fig:spec2c} compares our calculated
spectra B -- X  for the two isotopologues $^{26}$Al $^{16}$O  and $^{27}$Al
$^{16}$O. The shift in the band head feature should be observable astronomically
at even moderate resolution.

\begin{table*}
\caption{ Extract from the state file for $^{27}$Al$^{16}$O. Full tables
are available from
http://cdsarc.u-strasbg.fr/cgi-bin/VizieR?-source=J/MNRAS/xxx/yy.}
\label{tab:levels}
\begin{center}
\footnotesize
\tabcolsep=5pt
\begin{tabular}{rrrrrrlrrrr c rrr}
\hline
     $n$ & \multicolumn{1}{c}{$\tilde{E}$} &  $g$    & $J$  &  \multicolumn{1}{c}{$+/-$} &  \multicolumn{1}{c}{$e/f$} & State & $v$    & $|\Lambda|$& $|\Sigma|$ & $|\Omega|$& \\
\hline
           1  &       0.000000   &      12   &      0.5  &   +  &   e   &  \verb!X2SIGMA+ !&     0   &   0   &      0.5    &     0.5  \\
           2  &     965.435497   &      12   &      0.5  &   +  &   e   &  \verb!X2SIGMA+ !&     1   &   0   &      0.5    &     0.5  \\
           3  &    1916.845371   &      12   &      0.5  &   +  &   e   &  \verb!X2SIGMA+ !&     2   &   0   &      0.5    &     0.5  \\
           4  &    2854.206196   &      12   &      0.5  &   +  &   e   &  \verb!X2SIGMA+ !&     3   &   0   &      0.5    &     0.5  \\
           5  &    3777.503929   &      12   &      0.5  &   +  &   e   &  \verb!X2SIGMA+ !&     4   &   0   &      0.5    &     0.5  \\
           6  &    4686.660386   &      12   &      0.5  &   +  &   e   &  \verb!X2SIGMA+ !&     5   &   0   &      0.5    &     0.5  \\
           7  &    5346.116382   &      12   &      0.5  &   +  &   e   &  \verb!A2PI     !&     0   &   1   &      0.5    &     0.5  \\
           8  &    5581.906844   &      12   &      0.5  &   +  &   e   &  \verb!X2SIGMA+ !&     6   &   0   &      0.5    &     0.5  \\
           9  &    6066.934830   &      12   &      0.5  &   +  &   e   &  \verb!A2PI     !&     1   &   1   &      0.5    &     0.5  \\
          10  &    6463.039443   &      12   &      0.5  &   +  &   e   &  \verb!X2SIGMA+ !&     7   &   0   &      0.5    &     0.5  \\
          11  &    6778.997803   &      12   &      0.5  &   +  &   e   &  \verb!A2PI     !&     2   &   1   &      0.5    &     0.5  \\
          12  &    7329.427637   &      12   &      0.5  &   +  &   e   &  \verb!X2SIGMA+ !&     8   &   0   &      0.5    &     0.5  \\
          13  &    7483.145675   &      12   &      0.5  &   +  &   e   &  \verb!A2PI     !&     3   &   1   &      0.5    &     0.5  \\
          14  &    8159.170405   &      12   &      0.5  &   +  &   e   &  \verb!A2PI     !&     4   &   1   &      0.5    &     0.5  \\
          15  &    8201.467744   &      12   &      0.5  &   +  &   e   &  \verb!X2SIGMA+ !&     9   &   0   &      0.5    &     0.5  \\
          16  &    8857.266385   &      12   &      0.5  &   +  &   e   &  \verb!A2PI     !&     5   &   1   &      0.5    &     0.5  \\
          17  &    9029.150380   &      12   &      0.5  &   +  &   e   &  \verb!X2SIGMA+ !&    10   &   0   &      0.5    &     0.5  \\
          18  &    9535.195842   &      12   &      0.5  &   +  &   e   &  \verb!A2PI     !&     6   &   1   &      0.5    &     0.5  \\
          19  &    9854.882567   &      12   &      0.5  &   +  &   e   &  \verb!X2SIGMA+ !&    11   &   0   &      0.5    &     0.5  \\
          20  &   10204.019475   &      12   &      0.5  &   +  &   e   &  \verb!A2PI     !&     7   &   1   &      0.5    &     0.5  \\
          21  &   10667.668381   &      12   &      0.5  &   +  &   e   &  \verb!X2SIGMA+ !&    12   &   0   &      0.5    &     0.5  \\
          22  &   10864.560220   &      12   &      0.5  &   +  &   e   &  \verb!A2PI     !&     8   &   1   &      0.5    &     0.5  \\
          23  &   11464.897083   &      12   &      0.5  &   +  &   e   &  \verb!X2SIGMA+ !&    13   &   0   &      0.5    &     0.5  \\
          24  &   11519.212123   &      12   &      0.5  &   +  &   e   &  \verb!A2PI     !&     9   &   1   &      0.5    &     0.5  \\
          25  &   12156.974798   &      12   &      0.5  &   +  &   e   &  \verb!A2PI     !&    10   &   1   &      0.5    &     0.5  \\
          26  &   12257.694655   &      12   &      0.5  &   +  &   e   &  \verb!X2SIGMA+ !&    14   &   0   &      0.5    &     0.5  \\
          27  &   12793.671660   &      12   &      0.5  &   +  &   e   &  \verb!A2PI     !&    11   &   1   &      0.5    &     0.5  \\
          28  &   13030.412255   &      12   &      0.5  &   +  &   e   &  \verb!X2SIGMA+ !&    15   &   0   &      0.5    &     0.5  \\
          29  &   13421.583651   &      12   &      0.5  &   +  &   e   &  \verb!A2PI     !&    12   &   1   &      0.5    &     0.5  \\
          30  &   13790.933964   &      12   &      0.5  &   +  &   e   &  \verb!X2SIGMA+ !&    16   &   0   &      0.5    &     0.5  \\
\hline

\end{tabular}
\end{center}

\mbox{}\\
{\flushleft
$n$:   State counting number.     \\
$\tilde{E}$: State energy in \cm. \\
$J$: Total angular momentum quantum number.\\
$g$: State degeneracy.            \\
$+/-$:   Total parity. \\
$e/f$:   Rotationless-parity \citep{75BrHoHu.parity}. \\
$v$:   State vibrational quantum number. \\
$|\Lambda|$:   Absolute value of $\Lambda$ (projection of the electronic angular momentum). \\
$|\Sigma|$:   Absolute value of $\Sigma$ (projection of the electronic spin). \\
$|\Omega|$:   Absolute value of $\Omega=\Lambda+\Sigma$ (projection of the
total angular momentum).}

\end{table*}

\begin{table}
\caption{ Extracts from the transitions file for $^{27}$Al$^{16}$O.
 Full tables are available from
http://cdsarc.u-strasbg.fr/cgi-bin/VizieR?-source=J/MNRAS/xxx/yy. }
\label{tab:trans}
\begin{center}
\begin{tabular}{rrr}
\hline\hline
$f$	&  $i$ 		& 		$A_{fi}$\\
\hline
   47156     &     47355   &$  1.1598E-04  $   \\
    9373     &      8773   &$  5.0797E-02  $   \\
   10989     &     10389   &$  1.5734E-02  $   \\
   10789     &     10589   &$  1.5455E-02  $   \\
    9755     &      9155   &$  6.3206E-03  $   \\
   12788     &     13387   &$  2.7204E-06  $   \\
   10178     &      9578   &$  2.3282E-02  $   \\
    9555     &      9355   &$  6.7365E-03  $   \\
    9187     &      8987   &$  5.4996E-02  $   \\
    9587     &      9387   &$  5.4633E-02  $   \\
    7360     &      7159   &$  9.1545E-06  $   \\
    9184     &      9384   &$  4.8954E-05  $   \\
    9751     &      9151   &$  5.9125E-03  $   \\
    9551     &      9351   &$  5.9229E-03  $   \\
   10166     &      9566   &$  8.9220E-03  $   \\
   10985     &     10385   &$  1.4584E-02  $   \\
   10785     &     10585   &$  1.4584E-02  $   \\
    8548     &      7948   &$  5.1842E-02  $   \\
   20975     &     20775   &$  5.7037E-05  $   \\
    8148     &      7548   &$  5.2229E-02  $   \\
    9966     &      9766   &$  9.1712E-03  $   \\
\hline
\end{tabular}

\noindent
 $f$: Upper state counting number;
$i$:      Lower state counting number; $A_{fi}$:  Einstein-A coefficient in
s$^{-1}$.

\end{center}
\end{table}


\section{Conclusions}

We present comprehensive line lists for the four most important
isotopologues of AlO. These are based on the direct solution of the
nuclear motion Schr\"odinger equation using a potential energy curves and couplings
obtained by fitting to extensive dataset of measured transitions.
These data are reproduced to near experimental accuracy resulting in
high accuracy line positions. A new {\it ab initio} dipole moment is
computed. This dipole is used to compute Einstein A
coefficients for all possible dipole-allowed transitions within each
AlO isotopologue. The result is a comprehensive line list for each
species. The line
lists can be downloaded from the CDS, via
ftp://cdsarc.u-strasbg.fr/pub/cats/J/MNRAS/, or
http://cdsarc.u-strasbg.fr/viz-bin/qcat?J/MNRAS/, or from
www.exomol.com.

\section*{Acknowledgements}

This work is supported by ERC Advanced Investigator Project 267219.

\bibliographystyle{mn2e}

\begin{thebibliography}{44}
\expandafter\ifx\csname natexlab\endcsname\relax\def\natexlab#1{#1}\fi

\bibitem[{Bai {et~al}\mbox{.}({2014})Bai, Motto-Ros, Lei, Zheng, \&
  Yu}]{14BaMoLe.AlO}
Bai X., Motto-Ros V., Lei W., Zheng L., Yu J., {2014}, Spectra Chimica Acta B,
  {99}, 193

\bibitem[{Banerjee {et~al}\mbox{.}({2004})Banerjee, Ashok, Launila, Davis, \&
  Varricatt}]{04BaAsLa.AlO}
Banerjee D. P.~K., Ashok N.~M., Launila O., Davis C.~J., Varricatt W.~P.,
  {2004}, ApJ, {610}, L29

\bibitem[{Banerjee {et~al}\mbox{.}(2005)Banerjee, Barber, Ashok, \&
  Tennyson}]{jt357}
Banerjee D. P.~K., Barber R.~J., Ashok N.~K., Tennyson J., 2005, ApJ, 627, L141

\bibitem[{Banerjee {et~al}\mbox{.}(2012)Banerjee, Varricatt, Mathew, Launila,
  \& Ashok}]{12BaVaMa.AlO}
Banerjee D. P.~K., Varricatt W.~P., Mathew B., Launila O., Ashok N.~M., 2012,
  Astrophys. J. Lett., 753, L20

\bibitem[{Barber {et~al}\mbox{.}(2006)Barber, Tennyson, Harris, \&
  Tolchenov}]{jt378}
Barber R.~J., Tennyson J., Harris G.~J., Tolchenov R.~N., 2006, MNRAS, 368,
  1087

\bibitem[{Barton {et~al}\mbox{.}(2014)Barton, Chiu, Golpayegani, Yurchenko,
  Tennyson, Frohman, \& Bernath}]{jt583}
Barton E.~J., Chiu C., Golpayegani S., Yurchenko S.~N., Tennyson J., Frohman
  D.~J., Bernath P.~F., 2014, MNRAS, 442, 1821

\bibitem[{Barton {et~al}\mbox{.}(2013)Barton, Yurchenko, \& Tennyson}]{jt563}
Barton E.~J., Yurchenko S.~N., Tennyson J., 2013, MNRAS, 434, 1469

\bibitem[{Bernard \& Gravina(1984)}]{84BeGrxx.AlO}
Bernard A., Gravina R., 1984, Z. Naturfors. Sect. A-J. Phys. Sci., 39, 1049

\bibitem[{Bescos {et~al}\mbox{.}(1995)Bescos, Morley, \& Urena}]{95BeMoUr.AlO}
Bescos B., Morley G., Urena A.~G., 1995, Chem. Phys. Lett., 244, 407

\bibitem[{Brown {et~al}\mbox{.}(1975)Brown, Hougen, Huber, Johns, Kopp,
  Lefebvre-Brion, Merer, Ramsay, Rostas, \& Zare}]{75BrHoHu.parity}
Brown J.~M. {et~al.}, 1975, J. Mol. Spectrosc., 55, 500

\bibitem[{Coxon \& Naxakis(1985)}]{85CoNaxx.AlO}
Coxon J.~A., Naxakis S., 1985, J. Mol. Spectrosc., 111, 102

\bibitem[{Dagdigian {et~al}\mbox{.}(1975)Dagdigian, Cruse, \&
  Zare}]{75DaCrZa.AlO}
Dagdigian P.~J., Cruse H.~W., Zare R.~N., 1975, J. Chem. Phys., 62, 1824

\bibitem[{Fischer {et~al}\mbox{.}(2003)Fischer, Gamache, Goldman, Rothman, \&
  Perrin}]{03FiGaGo.partfunc}
Fischer J., Gamache R.~R., Goldman A., Rothman L.~S., Perrin A., 2003, J.
  Quant. Spectrosc. Radiat. Transf., 82, 401

\bibitem[{Glumac {et~al}\mbox{.}(2001)Glumac, Servaites, \&
  Krier}]{01GlSeKr.AlO}
Glumac N.~G., Servaites J., Krier H., 2001, Combust. Sci. Technol., 172, 97

\bibitem[{Hill {et~al}\mbox{.}(2013)Hill, Yurchenko, \& Tennyson}]{jt542}
Hill C., Yurchenko S.~N., Tennyson J., 2013, Icarus, 226, 1673

\bibitem[{Johnson(1965)}]{65Joxxxx.AlO}
Johnson E.~R., 1965, J. Geophys. Res., 70, 1275

\bibitem[{Johnson {et~al}\mbox{.}(1972)Johnson, Capelle, \&
  Broida}]{72JoCaBr.AlO}
Johnson S.~E., Capelle G., Broida H.~P., 1972, J. Chem. Phys., 56, 663

\bibitem[{Kaminski {et~al}\mbox{.}(2013)Kaminski, Schmidt, \&
  Menten}]{13KaScMe.AlO}
Kaminski T., Schmidt M.~R., Menten K.~M., 2013, A\&A, 549, A6

\bibitem[{Knecht {et~al}\mbox{.}({1996})Knecht, Pike, Murad, \&
  Rall}]{96KnPiMu.AlO}
Knecht D.~J., Pike C.~P., Murad E., Rall D. L.~A., {1996}, J. Spacecrafts
  Rockets, {33}, 677

\bibitem[{Launila \& Berg(2011)}]{11LaBexx.AlO}
Launila O., Berg L.-E., 2011, J. Mol. Spectrosc., 265, 10

\bibitem[{Launila \& Jonsson(1994)}]{94LaJoxx.AlO}
Launila O., Jonsson J., 1994, J. Mol. Spectrosc., 168, 1

\bibitem[{Lengsfield \& Liu(1982)}]{82LeLixx.AlO}
Lengsfield B.~H., Liu B., 1982, J. Chem. Phys., 77, 6083

\bibitem[{Merrill {et~al}\mbox{.}({1962})Merrill, Keenan, \&
  Deutsch}]{62MeKeDe.AlO}
Merrill P.~W., Keenan P.~C., Deutsch A.~J., {1962}, ApJ, {136}, 21

\bibitem[{Naulin \& Costes(1999)}]{99NaCoxx.AlO}
Naulin C., Costes M., 1999, Chem. Phys. Lett., 310, 231

\bibitem[{Parigger \& Hornkohl(2011)}]{11PaHoxx.AlO}
Parigger C.~G., Hornkohl J.~O., 2011, Spectra Chimica Acta A, 81, 404

\bibitem[{Partridge {et~al}\mbox{.}(1983)Partridge, Langhoff, Lengsfield, \&
  Liu}]{83PaLaLe.AlO}
Partridge H., Langhoff S.~R., Lengsfield B.~H., Liu B., 1983, J. Quant.
  Spectrosc. Radiat. Transf., 30, 449

\bibitem[{Patrascu {et~al}\mbox{.}(2014)Patrascu, Hill, Tennyson, \&
  Yurchenko}]{jt589}
Patrascu A.~T., Hill C., Tennyson J., Yurchenko S.~N., 2014, J. Chem. Phys.,
  141, 144312

\bibitem[{Pickett {et~al}\mbox{.}(1998)Pickett, Poynter, Cohen, Delitsky,
  Pearson, \& {M\"uller}}]{jpl}
Pickett H.~M., Poynter R.~L., Cohen E.~A., Delitsky M.~L., Pearson J.~C.,
  {M\"uller} H. S.~P., 1998, J. Quant. Spectrosc. Radiat. Transf., 60, 883

\bibitem[{Saksena {et~al}\mbox{.}(2008)Saksena, Deo, Sunanda, Behere, \&
  Londhe}]{08SaDeSu.AlO}
Saksena M.~D., Deo M.~N., Sunanda K., Behere S.~H., Londhe C.~T., 2008, J. Mol.
  Spectrosc., 247, 47

\bibitem[{Sauval \& Tatum(1984)}]{84SaTaxx.partfunc}
Sauval A.~J., Tatum J.~B., 1984, ApJS, 56, 193

\bibitem[{Sriramachandran {et~al}\mbox{.}(2013)Sriramachandran, Viswanathan, \&
  Shanmugavel}]{13SrViSh.AlO}
Sriramachandran P., Viswanathan B., Shanmugavel R., 2013, Sol. Phys., 286, 315

\bibitem[{Surmick \& Parigger({2014})}]{14SuPa.AlO}
Surmick D.~M., Parigger C.~G., {2014}, Appl. Spectrosc., {68}, 992

\bibitem[{Tenenbaum \& Ziurys(2009)}]{09TeZixx.AlO}
Tenenbaum E.~D., Ziurys L.~M., 2009, ApJ, 694, L59

\bibitem[{Tennyson(2014)}]{jt573}
Tennyson J., 2014, J. Mol. Spectrosc., 298, 1

\bibitem[{Tennyson {et~al}\mbox{.}(2013)Tennyson, Hill, \& Yurchenko}]{jt548}
Tennyson J., Hill C., Yurchenko S.~N., 2013, in AIP Conference Proceedings,
  Vol. 1545, 6$^{th}$ international conference on atomic and molecular data and
  their applications ICAMDATA-2012, AIP, New York, pp. 186--195

\bibitem[{Tennyson \& Yurchenko(2012)}]{jt528}
Tennyson J., Yurchenko S.~N., 2012, MNRAS, 425, 21

\bibitem[{Tylenda {et~al}\mbox{.}({2005})Tylenda, Crause, Gorny, \&
  Schmidt}]{05TyCrGo.AlO}
Tylenda R., Crause L.~A., Gorny S.~K., Schmidt M.~R., {2005}, A\&A, {439}, 651

\bibitem[{Vidler \& Tennyson(2000)}]{jt263}
Vidler M., Tennyson J., 2000, J. Chem. Phys., 113, 9766

\bibitem[{Werner {et~al}\mbox{.}(2010)Werner, Knowles, Lindh, Manby, \&
  {Sch\"{u}tz}}]{molpro.method}
Werner H.~J., Knowles P.~J., Lindh R., Manby F.~R., {Sch\"{u}tz} M., 2010,
  {MOLPRO}, a package of ab initio programs. See {\footnotesize
  \texttt{http://www.molpro.net/}}

\bibitem[{Yadin {et~al}\mbox{.}(2012)Yadin, Vaness, Conti, Hill, Yurchenko, \&
  Tennyson}]{jt529}
Yadin B., Vaness T., Conti P., Hill C., Yurchenko S.~N., Tennyson J., 2012,
  MNRAS, 425, 34

\bibitem[{Yorke {et~al}\mbox{.}(2014)Yorke, Yurchenko, Lodi, \&
  Tennyson}]{jt590}
Yorke L., Yurchenko S.~N., Lodi L., Tennyson J., 2014, MNRAS, 445, 1383

\bibitem[{Yurchenko {et~al}\mbox{.}(2015)Yurchenko, Lodi, Tennyson, \&
  Stolyarov}]{jt606}
Yurchenko S.~N., Lodi L., Tennyson J., Stolyarov A.~V., 2015, Comput. Phys.
  Commun.

\bibitem[{Zenouda {et~al}\mbox{.}(1999)Zenouda, Blottiau, Chambaud, \&
  Rosmus}]{99ZeBlCh.AlO}
Zenouda C., Blottiau P., Chambaud G., Rosmus P., 1999, J. Molec. Struct.
  (THEOCHEM), 458, 61

\bibitem[{Zhang \& Li(2003)}]{03ZhLixx.AlO}
Zhang S.~D., Li H.~Y., 2003, Chem. Res. Chin. Univ., 19, 320

\end{thebibliography}

\end{document}